\PassOptionsToPackage{table}{xcolor}
\documentclass[review]{elsarticle}

\usepackage{lineno,hyperref}
\usepackage{enumitem}
\usepackage{amsmath,amssymb,amsfonts,amsthm}
\usepackage[table]{xcolor}
\usepackage{subcaption}
\usepackage{tikz}
\usetikzlibrary{calc,chains,patterns,backgrounds,decorations,decorations.pathmorphing}

\newtheorem{thm}{Theorem}

\newtheorem{cor}[thm]{Corollary}
\newtheorem{prop}[thm]{Proposition}
\newdefinition{rmk}{Remark}
\newproof{pf}{Proof}
\newproof{pot}{Proof of Theorem \ref{thm2}}

\newcommand{\yell}{\cellcolor{gray!30}}
 \def\<#1>{\noindent$\boldsymbol{\langle}$ \textbf{\textit{\ignorespaces#1\unskip}} $\boldsymbol{\rangle}$}

\begin{document}

\begin{frontmatter}

\title{Results for the maximum weight planar subgraph problem}

\author{Diane Castonguay, Elis\^{a}ngela Silva Dias\footnote{Corresponding author.}, Leslie Richard Foulds}
\address{\{diane,elisangela,lesfoulds\}@inf.ufg.br\\
Instituto de Inform\'atica, Universidade Federal de Goi\'as\\
Alameda Palmeiras, Quadra D, Campus Samambaia\\
CEP 74001-970, Goi\^ania, Goi\'as, Brasil}

\begin{abstract}
The problem of finding the maximum-weight, planar subgraph of a finite, simple graph with nonnegative real edge weights is well known in industrial and electrical engineering, systems biology, sociology and finance. As the problem is known to be NP-hard, much research effort has been devoted over the years to attempt to improve a given approximate solution to the problem by using local moves applied to a planar embedding of the solution. It has long been established that any feasible solution to the problem, a maximal planar graph, can be transformed into any other (having the same vertex set) in a finite sequence of local moves of based on: (i) edge substitution and (ii) vertex relocation and it has been conjectured that moves of only type (i) are sufficient. In this note we settle this conjecture in the affirmative. Furthermore, contrary to recent supposition, we demonstrate that any maximal spanning tree of the original graph is not necessarily a part of any optimal solution to the problem. We hope these results will be useful in the design of future approximate methods for the problem.
\end{abstract}


\end{frontmatter}


\section{Introduction}
\label{sec:intro}

Many complex systems involve a finite, discrete set of objects and the relationships between them. In some cases it is natural to think of the system as a graph, with the objects represented by the vertices of the graph and the relationships by its edges. It is often useful to visualize the system via a drawing of the graph in the plane. However, when the system has a nontrivial number of objects and most (or all) pairs of objects are related, the corresponding graph drawing is often somewhat complex to visualize and interpret. What is traditionally done in this situation is to draw a graph that has all the objects (vertices) but only a strict subset of the relationships (edges). This has motivated researchers to pose the question of what is the best balance between relationship information and clarity of graph drawing? To make the concept of ``best'' meaningful, two assumptions are commonly made:
\begin{enumerate}[label=(\roman*)]
	\item there is exactly one relationship between pair of objects and the relationship is undirected, (symmetric) and
	\item there is a nonnegative, real-numbered weight (not necessarily finite) associated with each relationship.
\end{enumerate}

The challenge is then to construct a {\it spanning} graph drawing (containing all the vertices) and a subset of the edges with the highest weight (sum of edge weights) that strikes an acceptable balance between the preservation of relationship information and graph drawing clarity. One minimalist approach is to select a subset of the edges in such a way that the graph is a spanning tree, i.e., there is a unique path between each pair of vertices. Although spanning trees are easy to visualize, there is often a significant loss of system relationship information in tree drawing representations - a major disadvantage. Because the edge weights are assumed to be all nonnegative, it is usually desirable to draw a graph with the highest weight (to conserve system information) from among those that are  {\it planar} i.e., can be drawn with all edges intersecting only at the vertices (to promote visualization clarity). This induces the following, much-studied problem. Given a system that satisfies assumptions (i) and (ii) above, what is its planar spanning subgraph representation of highest weight? This problem is commonly formulated in terms of graph theory and is known as the maximum-weight planar subgraph problem (MWPSP). The MWPSP has some important applications: (i) as a subproblem of the plant (facility) layout problem in industrial engineering, where the vertices represent the activities of the facility and the edges the adjacencies between them in a plan of the facility \cite{Ahmadi-Ardestani-Foulds-Hojabri-Farahani-2015} (ii) integrated circuit design -- where the vertices are the electrical elements and the edges represent the physical connections between them \cite{Lengauer-2012} (iii) systems biology, where the vertices represent proteins and edges represent protein interactions in a metabolic network \cite{Song-Aste-DiMatteo-2007}, (iv) social system analysis -- where the vertices represent social agents (e.g. individuals, groups or companies) and edges represent social interaction \cite{Easley-Kleinberg-2010}, and (v) the filtering of data in correlation-based graphs in finance \cite{Tumminello-Aste-DiMatteo-Mantegna-2005}. Because of the very nature of applications (i) and (ii), any subgraph considered as a solution must essentially be planar, however solutions to (iii), (iv) and (v) need not necessarily be planar -- this property is merely a desirable asset to enhance visualization.

As the MWPSP is known to be strongly NP-hard \cite{Giffin-1984} it should come as no surprise that exact algorithms are capable of solving general MWPSP instances with only a relatively small number of vertices \cite{Ahmadi-Ardestani-Foulds-Hojabri-Farahani-2015, Junger-Mutzel-1996} and that reported research has focused mainly on approximate and heuristic methods and the improvement of the solutions they generate \cite{Massara-Matteo-Aste-2017}. The present note reports new results on solution properties and improvement strategies.

\section{The MWPSP problem}

The problem introduced in Section \ref{sec:intro} can be formally stated as an optimization problem in graph theory. For a fuller treatment of graph theory beyond the concepts introduced here, the reader is referred to \cite{Bondy-Murty-2008}. This note focuses on undirected, connected, simple graphs of the form $G = (V, E)$ with finite vertex set $V$ and edge set $E$, where $|V| = n$, $|E| = m$. Furthermore, it assumed that $G$ is nonnegatively edge-weighted in the sense that there exists a function $w: E\to \mathbb{R}^+$. Given such a graph $G$, the maximum-weight planar subgraph problem (MWPSP) involves searching the planar spanning subgraphs of $G$ to find the one with the highest sum of edge weights. Because the edge weights are all nonnegative, there will always be a solution that is a maximally planar subgraph of $G$ and from now on we confine the search for MWPSP solutions to these. It is well known that in any plane embedding of a maximal planar graph $G$ with $n\geq 3$ vertices, the boundaries of all the faces of G are 3-cycles.

\subsection{Transformational moves}

A given feasible solution to an MWPSP instance that is a maximal planar graph can be transformed
into another feasible solution that is also a maximal planar graph (having the same vertex set) by various
simple local topological moves. One such move is \textit{edge substitution}, denoted by $T_1$ in \cite{Massara-Matteo-Aste-2017}, where an edge is removed and replaced either by the other diagonal of the 4-cycle that its removal induces or, if the diagonal is already present, by a unique, new edge. Another transformational process is \textit{vertex relocation}, denoted by $T_3$ in \cite{Massara-Matteo-Aste-2017}, where a vertex of degree 3 (assuming one exists) is removed (creating a new face) and is located in another face.

It has been proved that any maximal planar graph can be transformed into any other maximal planar
graph with the same vertex set in a finite sequence of edge substitution and vertex relocation moves \cite{Foulds-Robinson-1979}. In this note we extend the result in \cite{Foulds-Robinson-1979} by proving that only edge substitution moves are sufficient to bring about the transformation.

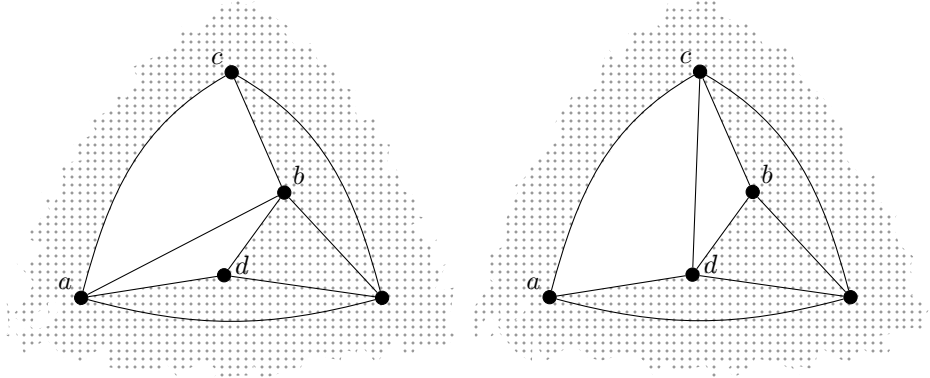
\begin{figure}[h!]
  \begin{subfigure}[t]{.45\textwidth}
   \centering
   \begin{tikzpicture}[
  	no/.style = {draw,circle,inner sep=1pt,minimum size=5pt,fill=black},
	]
    \node[no,label={[label distance =-3pt]105:$a$}] (a) at (0.0,0.0) {};
    \node[no] (p) at (4.0,0.0) {};
    \node[no,label={[label distance =-3pt]120:$c$}] (c) at (2.0,3.0) {};
    \node[no,label={[label distance =-3pt]45:$b$}] (b) at (2.7,1.4) {};
    \node[no,label={[label distance =-2pt,yshift=-3pt]3:$d$}] (d) at (1.9,0.3) {};
    \draw (a) to [out=-15,in=195] (p)
              (p) to [out=105,in=-30] (c)
              (c) to [out=210,in=75] (a);
    \draw (a) -- (d) -- (p) -- (b) -- (a);
    \draw (c) -- (b) -- (d);
    \begin{pgfonlayer}{background}
     \fill[pattern=dots, pattern color=black!40]
         (c.center) -- (b.center) -- (d.center) -- (a.center)
         decorate [decoration={random steps,segment length=4pt,amplitude=5pt}]
         {-- (-1.0,-0.5) to [out=-15,in=195] (5.0,-0.5) to [out=105,in=-30] (2.0,4.0) } -- cycle
         (c.center) to [out=210,in=75] (a.center)
         decorate [decoration={random steps,segment length=4pt,amplitude=5pt}]
         {-- (-1.0,-0.5) to [out=75,in=210] (2.0,4.0)} -- cycle;
    \end{pgfonlayer}
   \end{tikzpicture}
   \caption{Graph $G$ before the edge substitution move.}
   \label{sfig:ori-T1-1}
  \end{subfigure}\qquad
  \begin{subfigure}[t]{.45\textwidth}
   \centering
   \begin{tikzpicture}[
  	no/.style = {draw,circle,inner sep=1pt,minimum size=5pt,fill=black},
	]
    \node[no,label={[label distance =-3pt]105:$a$}] (a) at (0.0,0.0) {};
    \node[no] (p) at (4.0,0.0) {};
    \node[no,label={[label distance =-3pt]120:$c$}] (c) at (2.0,3.0) {};
    \node[no,label={[label distance =-3pt]45:$b$}] (b) at (2.7,1.4) {};
    \node[no,label={[label distance =-2pt,yshift=-3pt]3:$d$}] (d) at (1.9,0.3) {};
    \draw (a) to [out=-15,in=195] (p)
              (p) to [out=105,in=-30] (c)
              (c) to [out=210,in=75] (a);
    \draw (a) -- (d) -- (p) -- (b) -- (d) -- (c) -- (b);
    \begin{pgfonlayer}{background}
     \fill[pattern=dots, pattern color=black!40]
         (c.center) -- (b.center) -- (d.center) -- (a.center)
         decorate [decoration={random steps,segment length=4pt,amplitude=5pt}]
         {-- (-1.0,-0.5) to [out=-15,in=195] (5.0,-0.5) to [out=105,in=-30] (2.0,4.0) } -- cycle
         (c.center) to [out=210,in=75] (a.center)
         decorate [decoration={random steps,segment length=4pt,amplitude=5pt}]
         {-- (-1.0,-0.5) to [out=75,in=210] (2.0,4.0)} -- cycle;
    \end{pgfonlayer}
   \end{tikzpicture}
   \caption{The edge substitution move when $\{c, d\}$ in not an edge in $G$.}
   \label{sfig:new-T1-1}
  \end{subfigure}
  \caption{The edge substitution process (first case).}
  \label{fig:T1-move-1}
\end{figure}

The edge substitution and vertex relocation moves are now explained more formally. Let $G = (V, E, F)$ be a maximal planar graph and a planar embedding of it with vertex set $V$, edge set $E$, face set $F$ and $n\geq 5$. Let $f = |F|$. The edge substitution move is illustrated in Figure \ref{fig:T1-move-1} where $\{a, b, c\}, \{a, b, d\}\in F$ and $\{a, b\}\in E$ is the edge chosen to be substituted. Suppose $\{c, d\}\notin E$. The edge substitution move in this first case is illustrated in Figure \ref{fig:T1-move-1} where $\{a, b\}$ is replaced by $\{c,d\}$ to transform $G$ into the maximal planar graph $G' = (V', E', F')$ where $V' = V$, $E' = (E \setminus \{\{a, b\}\}) \cup \{\{c,
d\}\}$, $F' = (F \setminus \{\{a, b, c\}, \{a, b, d \}\}) \cup \{\{a, c, d\},\{b, c, d\}\}$. In this case the move involves substituting the edge $\{a, b\}$, of the diagonal of a 4-cycle $\langle a, c, b, d\rangle$, by the other (missing) diagonal, $\{c, d\}$.

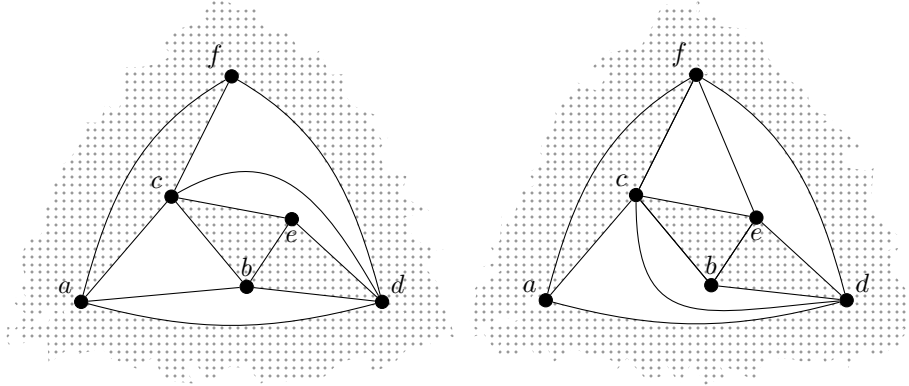
\begin{figure}[h!]
  \begin{subfigure}[t]{.45\textwidth}
   \centering
   \begin{tikzpicture}[
  	no/.style = {draw,circle,inner sep=1pt,minimum size=5pt,fill=black},
	]
    \node[no,label={[label distance =-3pt]105:$a$}] (a) at (0.0,0.0) {};
    \node[no,label={[label distance =-3pt]45:$d$}]  (d) at (4.0,0.0) {};
    \node[no,label={[label distance =-3pt]120:$f$}] (f) at (2.0,3.0) {};
    \node[no,label={[label distance =-2pt]92:$b$}] (b) at (2.2,0.20) {};
    \node[no,label={[label distance =-3pt]105:$c$}] (c) at (1.2,1.40) {};
    \node[no,label={[label distance =-2pt]-90:$e$}] (e) at (2.8,1.1) {};
    \draw (a) to [out=-15,in=195] (d)
              (d) to [out=105,in=-30] (f)
              (f) to [out=210,in=75] (a)
              (a) -- (b) -- (d) -- (e) -- (c) -- (a);
    \draw (f) -- (c) -- (b) -- (e);
    \draw (d) to [out=120,in=30,distance=1.5cm] (c);
   \begin{pgfonlayer}{background}
    \fill[pattern=dots, pattern color=black!40]
         (c.center) -- (b.center) -- (d.center) -- (e.center) -- cycle
         (f.center) -- (c.center) -- (a.center)
         decorate [decoration={random steps,segment length=4pt,amplitude=5pt}]
         {-- (-1.0,-0.5) to [out=75,in=210] (2.0,4.0)} -- cycle
         (d.center) to [out=105,in=-30] (f.center)
         decorate [decoration={random steps,segment length=4pt,amplitude=5pt}]
         {-- (2.0,4.0) to [out=-30,in=105] (5.0,-0.5)} -- cycle
         (d.center) to [out=195,in=-15] (a.center)
         decorate [decoration={random steps,segment length=4pt,amplitude=5pt}]
         {-- (-1.0,-0.5) to [out=-15,in=195] (5.0,-0.5)} -- cycle;
   \end{pgfonlayer}
   \end{tikzpicture}
   \caption{Graph $G$ before the edge substitution move.}
   \label{sfig:ori-T1-2}
  \end{subfigure}\qquad
  \begin{subfigure}[t]{.45\textwidth}
   \centering
   \begin{tikzpicture}[
  	no/.style = {draw,circle,inner sep=1pt,minimum size=5pt,fill=black},
	]
    \node[no,label={[label distance =-3pt]105:$a$}] (a) at (0.0,0.0) {};
    \node[no,label={[label distance =-3pt]45:$d$}]  (d) at (4.0,0.0) {};
    \node[no,label={[label distance =-3pt]120:$f$}] (f) at (2.0,3.0) {};
    \node[no,label={[label distance =-2pt]92:$b$}] (b) at (2.2,0.20) {};
    \node[no,label={[label distance =-3pt]105:$c$}] (c) at (1.2,1.40) {};
    \node[no,label={[label distance =-2pt]-90:$e$}] (e) at (2.8,1.1) {};
    \draw (a) to [out=-15,in=195] (d)
              (d) to [out=105,in=-30] (f)
              (f) to [out=210,in=75] (a)
              (a) -- (c) -- (e) -- (d) -- (b) -- (e) -- (f) -- (c) -- (b);
    \draw (f) -- (c) -- (b) -- (e);
    \draw (d) to [out=185,in=-90,distance=1.8cm] (c);
   \begin{pgfonlayer}{background}
    \fill[pattern=dots, pattern color=black!40]
         (c.center) -- (b.center) -- (d.center) -- (e.center) -- cycle
         (f.center) -- (c.center) -- (a.center)
         decorate [decoration={random steps,segment length=4pt,amplitude=5pt}]
         {-- (-1.0,-0.5) to [out=75,in=210] (2.0,4.0)} -- cycle
         (d.center) to [out=105,in=-30] (f.center)
         decorate [decoration={random steps,segment length=4pt,amplitude=5pt}]
         {-- (2.0,4.0) to [out=-30,in=105] (5.0,-0.5)} -- cycle
         (d.center) to [out=195,in=-15] (a.center)
         decorate [decoration={random steps,segment length=4pt,amplitude=5pt}]
         {-- (-1.0,-0.5) to [out=-15,in=195] (5.0,-0.5)} -- cycle;
   \end{pgfonlayer}
   \end{tikzpicture}
   \caption{The edge substitution move when $\{c, d\}$ in an edge in $G$.}
   \label{sfig:new-T1-2}
  \end{subfigure}
  \caption{The edge substitution process (second case).}
  \label{fig:T1-move-2}
\end{figure}

However, it is possible for both $\{a, b\}, \{c, d\}$ to be members of $E$. The edge substitution move in this second case is illustrated in Figure \ref{fig:T1-move-2}. Suppose faces $\{a, b, c\}, \{a, b, d\},$ $\{c, d, e\}, \{c, d, f\} \in F$. Either $a, b, e, f$ are all distinct or one of
$a, b$ is coincident with one of $e, f$. In either case, the edge substitution move replaces edge $\{a, b\}$ by the edge $\{e, f\}$. Note that $\{e, f\}\notin E$ otherwise $a, b, c, d, e, f$ are vertices of the (nonplanar) complete bipartite graph connecting $a, d, e$ with $b, c, f$. Replacing $\{a, b\}$ by $\{e, f\}$, transforms $G$ into the maximal planar graph $G' = (V', E', F')$ where $V' = V$, $E' = (E \setminus \{\{a, b\}\}) \cup \{\{e, f\}\}$, $F' = (F \setminus \{\{a, b, c\}, \{a, b, d\}, \{c, d, e\}, \{c, d, f\}\}) \cup \{\{a, c, d\},\{b, c, d\}, \{c, e, f\}, \{d, e, f\}\}$. Note that if $\{a, b\} = \{e, f\}$ then $n = 4$ and no transformation is possible.

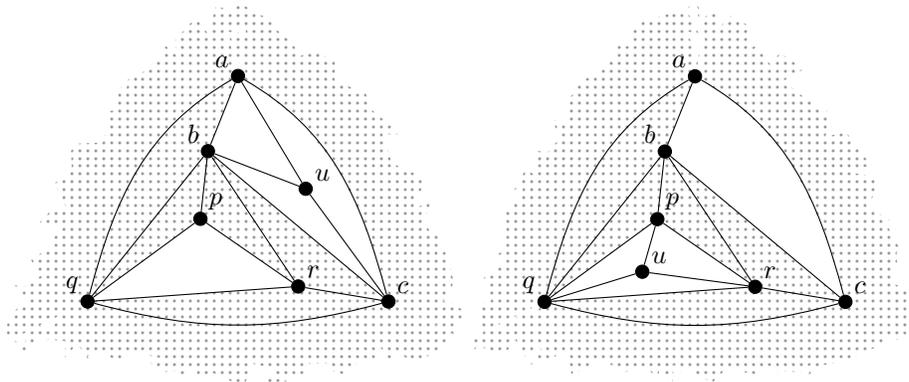
\begin{figure}[h!]
  \begin{subfigure}[t]{.45\textwidth}
   \centering
   \begin{tikzpicture}[
  	no/.style = {draw,circle,inner sep=1pt,minimum size=5pt,fill=black},
	]
    \node[no,label={[label distance =-3pt]105:$q$}] (q) at (0.0,0.0) {};
    \node[no,label={[label distance =-3pt]45:$c$}]  (c) at (4.0,0.0) {};
    \node[no,label={[label distance =-3pt]120:$a$}] (a) at (2.0,3.0) {};
    \node[no,label={[label distance =-3pt]105:$b$}] (b) at (1.6,2.0) {};
    \node[no,label={[label distance =-3pt]85:$u$}] (u) at (2.9,1.5) {};
    \node[no,label={[label distance =-3pt]5:$p$}] (p) at (1.5,1.1) {};
    \node[no,label={[label distance =-3pt]30:$r$}]  (r) at (2.8,0.2) {};
    \draw (q) to [out=-15,in=195] (c)
              (c) to [out=105,in=-30] (a)
              (a) to [out=210,in=75] (q);
    \draw (b) -- (a) -- (u) -- (b) -- (c) -- (u);
    \draw (c) -- (r) -- (b) -- (q) -- (r) -- (p) -- (q);
    \draw (b) -- (p);
   \begin{pgfonlayer}{background}
    \fill[pattern=dots, pattern color=black!40]
         (q.center) -- (p.center) -- (r.center) -- (c.center) -- (b.center) -- (a.center)
         decorate [decoration={random steps,segment length=4pt,amplitude=5pt}]
         {-- (2.0,4.0) to [out=210,in=75] (-1.0,-0.5)} -- cycle
         (c.center) -- (r.center) -- (q.center)
         decorate [decoration={random steps,segment length=4pt,amplitude=5pt}]
         {-- (-1.0,-0.5) to [out=-15,in=195] (5.0,-0.5)} -- cycle
         (c.center) to [out=105,in=-30] (a.center)
         decorate [decoration={random steps,segment length=4pt,amplitude=5pt}]
         {-- (2.0,4.0) to [out=-30,in=105] (5.0,-0.5)} -- cycle;
   \end{pgfonlayer}
   \end{tikzpicture}
   \caption{Graph $G$ before the vertex relocation move with vertex $u$ in face $\{a, b, c\}$.}
   \label{sfig:ori-T3}
  \end{subfigure}\qquad
  \begin{subfigure}[t]{.45\textwidth}
   \centering
   \begin{tikzpicture}[
  	no/.style = {draw,circle,inner sep=1pt,minimum size=5pt,fill=black},
	]
    \node[no,label={[label distance =-3pt]105:$q$}] (q) at (0.0,0.0) {};
    \node[no,label={[label distance =-3pt]45:$c$}]  (c) at (4.0,0.0) {};
    \node[no,label={[label distance =-3pt]120:$a$}] (a) at (2.0,3.0) {};
    \node[no,label={[label distance =-3pt]105:$b$}] (b) at (1.6,2.0) {};
    \node[no,label={[label distance =-3pt]85:$u$}] (u) at (1.3,0.4) {};
    \node[no,label={[label distance =-3pt]5:$p$}] (p) at (1.5,1.1) {};
    \node[no,label={[label distance =-3pt]30:$r$}]  (r) at (2.8,0.2) {};
    \draw (q) to [out=-15,in=195] (c)
              (c) to [out=105,in=-30] (a)
              (a) to [out=210,in=75] (q);
    \draw (a) -- (b) -- (q) -- (r) -- (c) -- (b) -- (r) -- (u) -- (q) -- (p) -- (u);
    \draw (b) -- (p) -- (r);
   \begin{pgfonlayer}{background}
    \fill[pattern=dots, pattern color=black!40]
         (q.center) -- (p.center) -- (r.center) -- (c.center) -- (b.center) -- (a.center)
         decorate [decoration={random steps,segment length=4pt,amplitude=5pt}]
         {-- (2.0,4.0) to [out=210,in=75] (-1.0,-0.5)} -- cycle
         (c.center) -- (r.center) -- (q.center)
         decorate [decoration={random steps,segment length=4pt,amplitude=5pt}]
         {-- (-1.0,-0.5) to [out=-15,in=195] (5.0,-0.5)} -- cycle
         (c.center) to [out=105,in=-30] (a.center)
         decorate [decoration={random steps,segment length=4pt,amplitude=5pt}]
         {-- (2.0,4.0) to [out=-30,in=105] (5.0,-0.5)} -- cycle;
   \end{pgfonlayer}
   \end{tikzpicture}
   \caption{The relocation of vertex $u$ in face $\{p, q, r\}$.}
   \label{sfig:new-T3}
  \end{subfigure}
  \caption{The vertex relocation move.}
  \label{fig:T3-move}
\end{figure}

The vertex relocation move is illustrated in Figure \ref{fig:T3-move}, where $G$ is a maximal planar graph with $n\geq 5$, having a
vertex $u$ of degree 3 and distinct faces $\{a, b, u\}, \{b, c, u\},$ $\{a, c, u\}, \{p, q, r\} \in F$. The vertex relocation move removes $u$, creating the face $\{a, b, c\}$ and inserts $u$ into the different face $\{p, q, r\}$ to transform $G$ into the maximal
planar graph $G' = (V', E', F')$ where $V' = V$, $E' = (E \setminus \{\{a, u\}, \{b, u\}, \{c, u\}\}) \cup \{\{p, u\}, \{q, u\}, \{r, u\}\}$, $F' =
(F \setminus \{\{a, b, u\}, \{b, c, u\},$ $\{a, c, u\}, \{p, q, r\}\}) \cup \{\{a, b, c\},$ $\{p, q, u\}, \{q, r, u\}, \{p, r, u\}\}$.

\section{Results and Discussion}

The next theorem, due to Foulds and Robinson \cite{Foulds-Robinson-1979}, shows the initial importance of edge substitution and vertex relocation.

\begin{thm} \label{theo:FR1979}
\cite{Foulds-Robinson-1979} If $G$ and $G'$ are maximal planar graphs with at least five vertices having the same vertex set, there exists a finite sequence of edge substitution and vertex relocation moves that transforms $G$ into $G'$.
\end{thm}

It was conjectured in \cite{Foulds-Robinson-1979} that the sequence of moves described in Theorem \ref{theo:FR1979} could always be
revised to consist of only edge substitution moves. One of our main results is to establish this as a simple corollary to Theorem \ref{teo:finite_sequence} below. In order to state the theorem we need the following notation. Let $G = (V, E, F)$ be a maximal planar graph with $n\geq 5$. If a edge substitution move is applied to edge $e \in E$, the transformed graph is denoted by $\alpha(G, e)$. If a vertex relocation move is applied to vertex $u \in V$, which is of degree 3 and is relocated in face $f \in F$, the
transformed graph is denoted by $\beta(G, u, f)$.

\begin{thm} \label{teo:finite_sequence}
Let $G = (V, E, F)$ be a maximal planar graph with $n\geq 5$, having at least one vertex of degree 3.
Let $u \in V$ be a vertex of degree 3 and $f \in F$. If $G' = \beta(G, u, f)$ then there exists a finite sequence of edge substitution moves that transforms $G$ into $G'$.
\end{thm}

\begin{pf}
Let $a, b, c \in V$ be the (only) three vertices adjacent to $u$. Consider $f' = \{a, b, u\} \in F$. By Proposition 10.9 in \cite{Bondy-Murty-2008} the dual graph of $G$ is connected. Thus, there exists a finite, positive integer $s$ such that $f = f_0, f_1 ,\ldots ,f_s
= f'$ such that $f_i$ and $f_{i+1} \in F$ are adjacent in the sense that they have an edge of $E$ in common, for $i = 0,\ldots, s-1$. We can assume that $u$ is not a vertex of $f$, otherwise $G = G'$. Furthermore, we assume that $u$ is not a
vertex of $f_{s-1}$ (otherwise set $f' = f_{s-1}$). Since $u$ is not a vertex of $f_{s-1}$, which, by assumption is adjacent to $f_s = f'$, we have $f_{s-1} = \{a, b, d\} \in F$ for some $d \in V$. Since $\{a, b, c\} \notin F$, it is clear that $d \neq c$ and $\{u, d\} \notin E$.
Let $G'' = \beta(G, u, f_{s-1}) = (V, E'', F'')$, say, $G_1 = \alpha(G, \{a, b\}) = (V, E_1 , F_1 )$, say, and $G_2 = \alpha(G_1, \{c, u\}) = (V,E_2, F_2)$, say. We shall prove that $G'' = G_2$. First, note that $G_2$ is well defined since $\{a, b, c\} \notin F$, and thus $\{u, c\} \in E_1$. Since $\{u, d\} \notin E$ it follows that $E_1 = (E \cup \{\{u, d\}\}) \setminus \{\{a, b\}\}$ and $F_1 = (F \cup \{\{a, d, u\}, \{b, d,u\}\}) \setminus \{\{a, b, u\}, \{a, b, d\}\}$. Clearly, $\{a, b\} \notin E_1$ but $\{a, c, u\}, \{b, c, u\} \in F_1$. Therefore $E_2 = (E_1 \cup \{\{a, b\}\}) \setminus \{\{c, u\}\} = (E \cup \{\{u, d\}\}) \setminus \{\{c, u\}\} = E''$ and $F_2 = (F_2 \cup \{\{a, b, c\}, \{a, b, u\}\}) \setminus \{\{a, c, u\}, \{b, c, u\}\} = (F \cup \{\{a, d, u\}, \{b, d, u\}, \{a, b, c\}\}) \setminus \{\{a, b, d\}, \{a, c, u\}, \{b, c, u\}\} = F''$.
The result follows by induction.
\end{pf}

The following corollary follows easily from Theorems 1 and 2 and shows that edge substitution moves are sufficient.

\begin{cor}\label{cor:finite_sequence}
If $G$ and $G'$ are maximal planar graphs of at least five vertices having the same vertex set, then
there exists a finite sequence of edge substitution moves that transforms $G$ into $G'$.
\end{cor}

Before discussing the implications of Corollary 3 for the MWPSP we present a further result.

\subsection*{A counterexample that an MST is not a necessary part of an optimal solution}
\label{ssec:counterexample}

Tumminello et al. \cite{Tumminello-Aste-DiMatteo-Mantegna-2005} presented an approximate construction algorithm for the MWPSP on a given edge-weighted graph $G$, that begins with the maximal spanning tree (MST) $T$ say, of $G$ of maximum total edge weight. The algorithm then progressively adds edges of $G$ to $T$ until a maximal planar subgraph of $G$ is produced, in which case the algorithm is terminated.

However, the following 8-vertex MWPSP instance $G_{CE}$, with edge weights in Table \ref{tab:counterexample} demonstrates that the MST is not a necessary part of an optimal solution to the MWPSP. The (unique) MST of $G_{CE}$ is the path $P_{CE} = (1, 2, 3, 4, 5, 6, 7, 8)$. An optimal solution to the MWPSP for $G_{CE}$ is the maximal planar subgraph with the edges in bold in Table \ref{tab:counterexample}, with weight $24$. Note that the edge $\{7, 8\}$ of the MST is not part of this solution. Indeed, any planar subgraph of GCS containing PCS has weight at most $23$.

One might surmise that is possible to find an optimal solution for the MWPSP problem that contains a maximal weighted subtree. The following proposition shows that this is not always the case.

\begin{prop}\label{prop:counterexample}
Let $G$ be the weighted graph given by incidence matrix in Table \ref{tab:counterexample}. No optimal solution for the MWPSP problem for $G$ contains the following unique maximal weighted subtree 
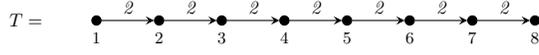
\begin{tikzpicture}[
    no/.style = {draw,circle,inner sep=1pt,minimum size=5pt,fill=black},
    >=stealth,
    start chain,
    transform shape,
    scale=.7
  ]
  \foreach \i in {1,...,8}
    \node[no,on chain,label={[font=\small]-90:$\i$}] (V\i) {};
  \node[draw=none,on chain,xshift=-11.2cm] (T) {$T=$};
  \foreach \j [count = \i] in {2,...,8}
   \draw[->] (V\i) -- node[above] {2} (V\j);
\end{tikzpicture}
\end{prop}

 \begin{table}[htbp!]
  \centering
  \begin{tabular}{cccccccc}
  \hline
  $w_{ij}$   & 2        & 3       & 4       & 5       & 6       & 7       & 8\\
  \hline
  1             &\yell \textbf{2} &\yell \textbf{1} &\yell \textbf{1} &\yell \textbf{1} & 0       & 0       &\yell \textbf{1}\\
  2             &          &\yell \textbf{2} & 0       &\yell \textbf{1} &\yell \textbf{1} & 0       &\yell \textbf{1}\\
  3             &          &          &\yell \textbf{2} & 0       &\yell \textbf{1} & 0       &\yell \textbf{1}\\
  4             &          &          &          &\yell \textbf{2} &\yell \textbf{1} &\yell \textbf{1} & 0\\
  5             &          &          &          &          &\yell \textbf{2} &\yell \textbf{1} & 0\\
  6             &          &          &          &          &          &\yell \textbf{2} & 0\\
  7             &          &          &          &          &          &          & 2\\
  \hline
  \end{tabular}
 \caption{Incidence matrix for graph $G$ of Proposition \ref{prop:counterexample}.}
 \label{tab:counterexample}
 \end{table}

We now discuss the implications of the above results. As has been previously stated, simple local
topological moves are often used to improve a feasible solution to an MWPSP instance \cite{Massara-Matteo-Aste-2017}. Corollary \ref{cor:finite_sequence} is an important aid in limiting the scope of such a strategy. Suppose graph G say, is a given feasible solution to an MWPSP instance and graph G' say, is an optimal solution. Corollary \ref{cor:finite_sequence} implies that there is always a way to transform G into G' using only a finite sequence edge substitution moves. Thus, only moves of edge substitution need be considered in the quest to transform a given feasible solution into an optimal one by applying only local topological moves.

Some construction algorithms for the MWPSP begin with a maximal spanning (MST) and extend it by adding edges to it until it is maximally planar \cite{Tumminello-Aste-DiMatteo-Mantegna-2005}. By Proposition \ref{prop:counterexample}, these do not necessarily lead to an optimal solution.

\section*{Acknowledgement}

The authors are grateful to Professor Humberto Longo of Federal University of Goi\'as, Brazil, 
for providing the figures in this note.


\bibliography{DEL-MWPSP}

\end{document}